\documentclass[12pt]{article}

\newif\ifblind
\blindfalse

\usepackage[authoryear]{natbib}
\usepackage{times}
\usepackage{setspace}
\usepackage{geometry}
\usepackage{amsmath,amssymb,amsfonts,mathtools}
\usepackage{booktabs}
\usepackage{array}
\usepackage{tabularx}
\usepackage{enumitem}
\usepackage{graphicx}
\usepackage{subcaption}
\usepackage{caption}
\usepackage{float}
\usepackage{tikz}
\usetikzlibrary{arrows.meta,positioning,fit,calc}
\usepackage{hyperref}
\usepackage{xcolor}
\usepackage{soul}

\newcommand{\E}{\mathbb{E}}

\newcommand{\1}{\mathbf{1}}
\newcommand{\bM}{\mathbf{M}}

\newcommand{\bphi}{\boldsymbol{\phi}}

\newcommand{\RR}{\mathbb{R}}

\newcommand{\BrokenRCTB}{1000}
\newcommand{\BegumN}{55}
\newcommand{\BegumTreat}{26}

\newcommand{\BegumTreatSharePct}{47.3\%}

\newcommand{\BegumXgbP}{0.0050}

\title{\Large
  Remote Auditing: Design-based Tests of Randomization, Selection, and Missingness with Broadly Accessible Satellite Imagery 
}

\ifblind
\author{\vspace{-0.5em}Anonymous Author(s)}
\date{}
\else
\author{
Connor T.\ Jerzak -- \textit{UT Austin, AI \& Global Development Lab}
\\ \;\;\;\; Adel Daoud -- \textit{Link\"{o}ping University, AI \& Global Development Lab}
}
\date{}
\fi

\begin{document}
\maketitle

\begin{abstract}
\noindent
\noindent Randomized controlled trials (RCTs) are the benchmark for causal inference, yet field implementation can drift from the registered design or, by chance, yield imbalances. We introduce a remote audit---a preregistrable, design‑based diagnostic that uses strictly pre‑treatment, publicly available satellite imagery to test whether assignment is independent of local conditions. The audit implements a conditional randomization test that asks whether treatment is more predictable from pre‑treatment features than under the registered mechanism, delivering a finite‑sample‑valid, nonparametric check that honors blocks and clusters and controls multiplicity across image models, resolutions, and patch sizes via a max‑statistic. The same preregistered procedure can be run before baseline data collection to guide implementation and, after assignments are realized, to audit the actual allocation. In two illustrations---Uganda’s Youth Opportunities Program (randomization corroborated) and a school‑based experiment in Bangladesh (assignment predictable relative to the design, consistent with independent concerns)---the audit can surface potential problems early, before costly scientific investments. We also provide descriptive diagnostics for selection into the study and for missingness. Because it is low‑cost and can be implemented rapidly in a unified way across diverse global administrative jurisdictions, the remote audit complements balance tests, strengthens preregistration, and enables rapid design checks when conventional data collection is slow, expensive, or infeasible.


\noindent \textbf{Word count:} 6,057
\end{abstract}

\newpage 
\singlespacing


\noindent Randomized experiments have transformed empirical social science by offering credible causal leverage in hard-to-study environments \citep{rubin2005causal,baldassarri2017field,gerber2017field}. In practice, however, the path from a pre-registered randomization mechanism to realized treatment assignment is sometimes fraught. Geography, logistics, bureaucratic discretion, political pressures, and bad randomization draws can all perturb assignment away from the intended design or away from covariate balance between treatment groups \citep{glennerster2013running,olken2015promises}. Even modest deviations can matter for inference, particularly when assignment correlates with contextual features that also shape outcomes \citep{battisti2017field}. Traditional safeguards such as centralized (re)randomization draws, sealed lists, enumerator training, together with ex-post diagnostics on covariate balance, manipulation checks, are indispensable \citep{morgan2012rerandomization,bruhn2009pursuit}. Yet they can be expensive, delayed, or underpowered in the very settings where field experiments are most valuable: low-resource environments, multi-jurisdiction programs, or government deployments in which baseline surveys are difficult or expensive to field at scale \citep{bouguen2019using}.

We here propose a complementary tool for field experiments: a \emph{remote audit} of randomization integrity that relies on pre-treatment satellite imagery.\footnote{Our scope is field experiments implemented in real-world settings; we do not target laboratory studies or small within-organization trials (e.g., within-school classrooms).} The idea is simple. Under the registered mechanism, treatment assignment should \textit{not} be predictable from covariates extracted from images collected \emph{before} randomization. If implementers implicitly targeted more accessible, wealthier, less conflict-prone, or otherwise distinctive places---attributes that often leave visual traces even at moderate resolution---then a predictive signal should be detectable in the imagery.

Our approach operationalizes these ideas as a \emph{conditional randomization test} (CRT) \citep{candes2018panning,hennessy2016conditional} tailored to field experiments. We (i) extract features from strictly pre-treatment images (e.g., Landsat, Sentinel) using interpretable indices (e.g., nightlight) or off-the-shelf backbones (e.g., CLIP-like encoders, ViT, Swin; \citet{xiao2025foundation}), (ii) train a predictive model of treatment using only these pre-treatment embeddings, summarize fit with an out-of-sample log-likelihood improvement statistic, and (iii) compare the observed statistic to its finite-sample reference distribution obtained by resampling from the known randomization scheme (honoring blocks, clusters, and treatment fractions). We further provide a max-statistic procedure \citep{westfall1993resampling} to adjust inference across multiple image models, resolutions, and patch sizes, and we discuss simple alternatives like Bonferroni or BH-style control \citep{thissen2002quick}. 

The remote audit of randomization quality is \emph{design-based}: validity does not hinge on outcome models or parametric assumptions, only on resampling from the registered assignment mechanism. It uses no post‑treatment variables, mitigating ``bad control'' risks \citep{angrist2009mostly,pearl2009causality}; and because we never condition on image features in the outcome model, the audit sidesteps adjustment‑based mediator and collider concerns. Pre‑treatment measurement helps but is not, by itself, sufficient for universally valid adjustment---see \citet{cinelli2020making}---which is precisely why we use imagery only to form a design‑based test of assignment. The audit complements standard balance tests: whereas balance checks examine low-dimensional covariates (often unavailable ex ante), the audit leverages high-dimensional, pre-existing visual context that is ubiquitous and pre-treatment.

The remote audit can be used both (i) ex ante (before or during field mobilization) to probe fidelity of implementation to stated design and guide remedial steps if issues are found, and (ii) ex post to evaluate the quality of a realized assignment vector. In this note, we focus on the pre‑treatment, randomization audit. We later describe two descriptive extensions---selection and missingness diagnostics---that can be run either ex ante or ex post, but are not design-valid tests in the same sense as for randomization. Operationally, these extensions retarget the same predictability exercise, asking whether pre‑treatment imagery predicts these labels better than a baseline. Because there is typically no registered mechanism for missingness or selection, we treat the resulting evidence as descriptive diagnostics rather than design‑valid tests, but the data, folds, and estimation machinery are identical to the randomization audit.

We illustrate with two audits. First, re-analyzing Uganda's government-run Youth Opportunities Program (YOP) RCT \citep{blattman2014generating} using only pre-2008 imagery, we find the observed assignment is no more predictable than resamples under the reported lottery, consistent with proper randomization. The same workflow highlights (i) strong predictability of trial participation relative to a national frame and (ii) image-predictive missingness, flagging external validity and data-loss risks. Second, for a school-based RCT in Bangladesh \citep{begum2022parental}, cluster assignment of treatment is itself highly predictable from pre-treatment features relative to the reported design---evidence consistent with independent concerns about irregularities \citep{Bonander2025}. These low-cost audits can shape fieldwork priorities, measurement strategies, and pre-analysis plans.

Our contribution is threefold. \textit{First}, we formalize a preregistrable conditional randomization test for randomization integrity explicitly adapted to experiments that leverage freely and globally available satellite imagery. \textit{Second}, we provide a practical workflow---pre‑treatment image selection, patching and scale, out‑of‑sample evaluation, and multiplicity control---and release a no‑code application that implements the audit at scale.\footnote{URL: \url{https://audit.planetarycausalinference.org}} \textit{Third}, we show empirically that remote audits are informative when baseline covariates are unavailable or delayed, illustrating randomization, selection, and missingness diagnostics in two prominent field experiments \citep{dreher2015aid,benyishay2022economic,weisberg2009total}.

Although remote audits extend the scope of checks and balances to increase the quality of field experiments, they will not detect all implementation problems. Many political and social processes are not visible from space, and clouds, revisit cycles, and spatial resolution impose constraints. Nonetheless, as a minimally invasive, design-based check, a remote audit can serve as an early warning system, bringing to light potential risks and guiding remedial steps before costly and slow downstream scientific investments. 

\section{The Remote Audit in Context: Related Work}

Researchers already use satellite imagery to measure outcomes, build covariates, and study heterogeneity \citep{jean2016combining,yeh2020using,jejoda2022_hetero,torres2022beyond}. Imagery has also been used in a model-based approach to mitigate confounding when pre-treatment signals proxy latent conditions \citep{sanford2021using,burke_using_2021}. Our contribution is orthogonal to these uses. We deploy imagery in a \emph{design-based} capacity: a conditional randomization test that asks whether realized treatment assignment is independent of satellite-derived features under the registered mechanism. Because the audit never adjusts outcomes, it avoids ``bad control'' pitfalls and mediator/collider concerns that attend post-treatment measurement in model-based pipelines \citep{angrist2009mostly,pearl2009causality,cinelli2020making}. Where model-based approaches typically require identification assumptions and sensitivity analysis \citep{rosenbaum1983assessing,oster2019unobservable}, our audit derives validity from the design itself: extremeness is judged against the experiment's finite-sample reference generated by draws from the assignment mechanism. In this sense, the paper reframes what imagery is---not as a control set within an outcome model, but as a ubiquitous, pre-existing source of design-stage information capable of certifying whether an assignment behaves as if it is random based on observables.

\section{A Conditional Randomization Test for Remote Audits}

\subsection{Design \& Data: Units, Imagery, Representations} 

Let \(\Omega\) denote the registered randomization procedure (e.g., complete randomization at rate \(\bar{a}\), or stratified randomization within blocks with fixed treatment counts). Consider experimental units \(i=1,\ldots,n\) (e.g., villages, neighborhoods, clinics), each with geospatial coordinates \(\ell_i \in \RR^2\). Let \(A_i\in\{0,1\}\) denote treatment assignment and \(Y_i\) the outcome. Prior to any intervention, we extract a pre-treatment image patch of size \(s>0\) centered on \(\ell_i\) from a public image archive (e.g., Landsat), denoted \(\bM_i = f_M(\ell_i,s)\). 

From $\bM_i$, we compute a representation $\bphi_i=f_{\bphi}(\bM_i)\in\RR^d$ using either interpretable indices (e.g., vegetation or texture) or off‑the‑shelf encoders (e.g., CLIP‑like, ViT, Swin), or both. Here, by \emph{representation} we mean a fixed‑length numeric summary of an image that compresses visible patterns such as roads, settlement structure, vegetation, and roof materials into measurements usable downstream by standard predictive models.\footnote{Embeddings are \emph{generated variables}. When used as regressors for causal estimation, they can require additional care \citep{battaglia2024inference}. Our design-based test avoids such complications because we only use embeddings to form a test statistic for assignment, not to adjust outcome models.} We use a representation $\bphi_i=f_{\bphi}(\bM_i)$ rather than the raw image $\bM_i$ to make evaluation computationally tractable; similar logic applies to both.

Under the experiment’s intended design $\Omega$, assignment is independent of all pre-treatment variables, including any fixed representation of the image:
\[
A_i \;\perp\!\!\!\perp\; \bM_i \quad\Rightarrow\quad A_i \;\perp\!\!\!\perp\; \bphi_i ,
\]
which is equivalent to the statement that, conditional on $\Omega$, $\bphi_i$ contains no information for predicting $A_i$ beyond the baseline treatment probabilities implied by the design.\footnote{If blocking or stratification is used, independence should hold within clusters, not unconditionally.}


 For \emph{validity}, we require only that $f_{\bphi}$ is fixed ex ante and that $\bphi_i$ is constructed strictly \emph{pre‑treatment} from $\bM_i$ captured before any mobilization. Under $\Omega$, $A_i\perp\!\!\!\perp\bM_i$ implies $A_i\perp\!\!\!\perp\bphi_i$ for any fixed $f_{\bphi}$, so CRT $p$‑values are finite‑sample valid regardless of how informative $\bphi_i$ is. For \emph{power}, richer $\bphi_i$ (e.g., pretrained encoders plus interpretable indices) help detect departures when they exist, but discarding information can only reduce power, not invalidate the test. 


Informally, \(\bphi_i\) should not help predict \(A_i\) beyond the treatment fraction implied by \(\Omega\). If \(\bphi_i\) \emph{does} predict \(A_i\) in the observed data substantially better than it typically does under draws from \(\Omega\), that is evidence that the realized assignment is atypical of the design---consistent with implementer discretion, operational constraints, or administrative errors aligning treatment with visual correlates of local conditions. In contrast to covariate balance tests of experimentor-collected features, which examine a low-dimensional set of pre-specified variables, remote audits can leverage high-dimensional pre-treatment information available almost everywhere on Earth.\footnote{Earth-observation missions provide global coverage and long archives, though revisit, cloud cover, and licensing constraints matter \citep{barnum2022dealing,tao2016did,townsend2021remote}.} 

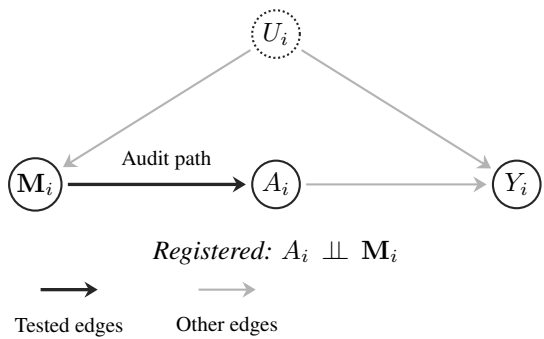
\begin{figure}[htb]
\centering
\begin{tikzpicture}[
  >=Stealth, font=\footnotesize, shorten >=2pt,shorten <=2pt
]
  \definecolor{ink}{gray}{0.15}
  \definecolor{muted}{gray}{0.70}
  \tikzset{
    var/.style   ={circle,draw=ink, line width=0.8pt, inner sep=1.2pt, minimum size=18pt},
    latent/.style={var, densely dotted},
    edgeBase/.style  ={-{Stealth[length=5pt,width=6pt]}, line width=0.8pt, draw=muted},
    edgeAudit/.style ={-{Stealth[length=5pt,width=6pt]}, line width=1.25pt, draw=ink},
    note/.style      ={font=\footnotesize, align=center}
  }
  \node[var]   (M)   at (0,0)         {$\bM_i$};
  \node[var]   (A)   at (3.2,0)       {$A_i$};
  \node[var]   (Y)   at (6.4,0)       {$Y_i$};
  \node[latent](U)   at (3.2,2.0)     {$U_i$};
  \draw[edgeBase] (A) -- (Y);
  \draw[edgeBase] (U) -- (Y);
  \draw[edgeBase] (U) -- (M);
  \draw[edgeAudit] (M) -- node[note,above,pos=0.55,yshift=1pt] {\scriptsize Audit path} (A);
  \node[note] at (3.2,-0.9) {\(\textit{Registered: } A_i \,\perp\!\!\!\perp\, \bM_i\)};
  \draw[edgeAudit] (0.0,-1.4) -- ++(0.9,0) node[note,below=1pt,xshift=-0.45cm,yshift=-0.9ex] {\scriptsize Tested edges};
  \draw[edgeBase]  (2.1,-1.4) -- ++(0.9,0) node[note,below=1pt,xshift=-0.45cm,yshift=-0.9ex] {\scriptsize Other edges};
\end{tikzpicture}
\caption{\textit{Remote audit intuition.} Is latent context \(U_i\), proxied by pre‑treatment imagery \(\bM_i\), correlated with treatment \(A_i\)? Under the registered mechanism, there is no \(\bM_i\!\to\!A_i\) edge; the CRT probes whether the realized assignment behaves as if such an edge were present.}
\label{fig:dag}
\end{figure}

Figure~\ref{fig:dag} encodes the estimand system at the level of raw pre‑treatment imagery: $\bM_i$ (what satellites see before any intervention), assignment $A_i$, outcome $Y_i$, and latent context $U_i$ that shapes both what satellites see and potential outcomes. Under the registered mechanism there is \emph{no} edge $\bM_i\!\to\!A_i$, i.e., $A_i \perp\!\!\!\perp \bM_i$. The audit asks whether the realized data behave as if an effective $\bM_i\!\to\!A_i$ link were present. Substantively, such a link could arise if implementers followed a \emph{human map} or administrative rule that favors, say, road-adjacent or wealthier-looking areas---patterns that $\bM_i$ proxies even if those rules never referenced imagery explicitly. We subsequently use features or embeddings, $\bphi_i=f_{\bphi}(\bM_i)$, purely as a computational device to test for such a link.



\subsection{Test Statistic: Out-of-Sample Likelihood Improvement}

\noindent To turn this intuition into a test, we need a single, learner-agnostic statistic that captures how much the pre-treatment image information predicts treatment assignment \emph{beyond} what the registered mechanism \(\Omega\) would yield by chance. The desiderata are simple: evaluate strictly out of sample to prevent overfitting; anchor the scale to \(\Omega\) so that ``no signal'' maps asymptotically to zero; allow additivity across units and folds for transparent cross-fitting; and avoid dependence on any particular classifier.

A natural choice satisfying these criteria is the improvement in the predictive log-likelihood of \(A\) on held-out data relative to the baseline assignment probabilities implied by \(\Omega\). Split the sample (or use $K$-fold cross-fitting). Fit a predictive model for $A_i$ using only $\bphi_i$ on the training fold; let $\hat\pi_i=\widehat{\Pr}(A_i=1\mid \bphi_i)$ be its predictions on the test fold. Define:
\[
\mathcal{L} \;=\; \sum_{i\in\text{test}} \left( A_i \log \hat\pi_i + (1-A_i)\log(1-\hat\pi_i)\right),
\]
and let $\bar a$ be the marginal treatment rate under $\Omega$. Our test statistic is:
\begin{equation}
(T=) \; \Delta\mathcal{L} \;\equiv\; \mathcal{L} \;-\; \sum_{i\in\text{test}} \big( A_i \log \bar{a} + (1-A_i)\log(1-\bar{a})\big).
\label{eq:stat}
\end{equation}
Under genuine randomization, strictly out-of-sample fitting makes \( \Delta\mathcal{L} \) concentrate near zero. In fact, when there is no signal, the baseline assignment probabilities implied by \( \Omega \) (the global rate \( \bar a \) are Bayes optimal: they maximize the \emph{expected} held-out log-likelihood. Any learner that perturbs these baselines without real information will, by the concavity of \( \log \), on average \emph{lose} log-likelihood on test folds. Hence, \( \E_{\Omega}[\Delta\mathcal{L}] \le 0 \) with equality only when \( \hat\pi_i \) collapses to the baselines. Small negative values of \(\Delta\mathcal{L}\) are therefore common in finite samples due to training noise and cross-fitting variability. Substantially positive values, by contrast, require genuine predictability from pre-treatment information; the Appendix details the finite-sample randomization reference under \( \Omega \) used to quantify extremeness.

In more complex designs, treatment assignments may vary for each unit (e.g., within blocks), the baseline log-likelihood improvement can be rewritten using unit baseline assignment probabilities $q_i \;\equiv\; \Pr_{A\sim\Omega}(A_i=1)$: 
\begin{equation}
(T=) \;\Delta\mathcal{L}^{\star} \;\equiv\; 
\underbrace{\sum_{i\in\text{test}} \Big[ A_i\log\hat\pi_i + (1-A_i)\log(1-\hat\pi_i) \Big]}_{\mathcal{L}}
\;-\;
\underbrace{\sum_{i\in\text{test}} \Big[ A_i\log q_i + (1-A_i)\log(1-q_i)\Big]}_{\mathcal{L}_0^*}.
\label{eq:stat_qi}
\end{equation}
The ``no-signal'' anchor is here again \(0\); in finite samples, values below 0 are, as just discussed, not uncommon. The same construction underlies the auxiliary audits, swapping the treatment with missingness and selection into study indicators.

The log-likelihood improvements in Eq. \ref{eq:stat} and Eq. \ref{eq:stat_qi} are instances of strictly proper scoring rules \citep{gneiting2007strictly}, ensuring that increases in \(T\) correspond to genuine predictive gains on the test fold (which are assessed against the randomization reference). Alternative proper scores (e.g., Brier improvement, defined as the mean squared error of the predicted probability versus the outcome) are admissible; we use the likelihood scale here because it accumulates naturally across folds and aligns with design-based resampling.

\subsection{Reference Distribution and $p$-Values}

To interpret the statistic \(T\), we need a reference distribution that encodes the experiment's null: under the registered mechanism \(\Omega\), assignment is independent of all pre-treatment information.

The cleanest benchmark is randomization-based. We redraw assignment vectors from \(\Omega\) while holding fixed everything that is genuinely pre-treatment---the imagery, the derived embeddings, all preprocessing choices, and the train/test split used for cross-fitting---and for each redraw we refit the learner on the training fold and evaluate on the test fold to recompute the same likelihood-improvement statistic. Because the resampling respects block and cluster constraints by construction, the resulting reference distribution is calibrated in finite samples and requires no parametric approximation. Most importantly, the realized statistic and its resampled counterparts are exchangeable, so the rank of the observed \(T\) among them yields a valid measure of extremeness (a \(p\)-value). This is the essence of the conditional randomization test: a design-based calibration that turns high-dimensional predictability into evidence about departures from the intended randomization \citep{candes2018panning,hennessy2016conditional}. We now make this construction explicit.

For \(b=1,\dots,B\), resample treatment vectors \(A^{(b)}\sim \Omega\) subject to the same constraints (block sizes, treatment quotas), re-fit the predictive model on the training fold with \((\bphi, A^{(b)})\), re-compute \(\hat\pi^{(b)}\) on the test fold, and compute \(T^{(b)}\) exactly as above. The CRT $p$-value is
\[
p \;=\; \frac{1}{B+1}\left(1+\sum_{b=1}^B \1\{T^{(b)} \ge T\}\right),
\]
which is valid in finite samples for arbitrary test statistics provided the resampling respects \(\Omega\) \citep{candes2018panning,hennessy2016conditional}. This procedure is nonparametric, transparently honors blocking/stratification, and accommodates any off-the-shelf learner as the engine of the statistic. See Appendix for a simple sketch of validity.

An important implementation detail concerns re-fitting versus re-using the trained learner while remote auditing. For strict finite-sample validity, we must refit within each resample using the same cross-fitting protocol (assuming the learner depends on the labels). In moderate samples with simple learners and simple image features, such as NDVI (a measure of vegetation), the computational burden is modest. 

That said, when the remote audit compresses pre-treatment imagery into high-dimensional embeddings $\bphi_i\in\mathbb{R}^d$ (with $d$ often in the hundreds or thousands, arising from off-the-shelf encoders), the primary computational burden is re-fitting learners within each resample while preserving the design-based calibration to $\Omega$. We can therefore precompute $\bphi$ once and apply variance-reduction devices that \emph{do not} alter the reference distribution. For example, one variance reduction device that can reduce the number of Monte Carlo iterations needed is ``common random numbers'' \citep{wright1979effectiveness}: fix the cross-fitting split and all model-training seeds across resamples so that variation in $T^{(b)}$ arises solely from the assignment draws $A^{(b)}\sim\Omega$. This keeps the statistic exchangeable with its resampled counterparts while achieving target Monte Carlo precision with fewer iterations. Second, in designs that treat exactly half of the units within each block, pairing every draw with its blockwise complement implements an antithetic coupling that further reduces Monte Carlo variance, provided the complement map preserves $\Omega$. Finally, parallelization across $b$ is straightforward. These choices keep the high-dimensional, imagery-driven audit inexpensive and fast without diluting its guarantees.



What about multiple testing in the remote audit? If several embeddings or hyperparameters are considered (e.g., CLIP-like encoders, ViT, Swin; patch sizes; resolutions), remote audits should control the family-wise error rate (FWER) or the false discovery rate (FDR). A simple and powerful approach for FWER is the \emph{Westfall–Young max-$T$} correction \citep{westfall1993resampling}: at each resampled assignment \(A^{(b)}\), compute \emph{every} model's \(T^{(b)}\) and record the maximum. Compare the observed \(T\) for each model to this max distribution to obtain adjusted $p$-values. Alternatively, Bonferroni or Benjamini–Hochberg-style methods are available \citep{thissen2002quick}. We recommend preregistering the model set and correction rule if used.

Putting the test statistic and its reference distribution together, power will be higher when deviations from the registered mechanism align with pre-treatment visual signals and when the embedding/learner captures stable structure under out-of-sample evaluation. If assignment turns on factors that satellites cannot see (e.g., patronage networks), the audit will have limited power. By contrast, when assignment covaries with roads, settlement density, roof materials, or land cover---features that proxy accessibility and wealth and are reliably visible from space \citep{henderson2012measuring,jean2016combining,watmough_socioecologically_2019,burke_using_2021}---the test will be more informative.


\paragraph{Auxiliary audits.} As noted, similar logic supports audits of (i) \emph{selection into the experiment} (predicting membership among a broader frame) and (ii) \emph{missingness} (predicting which units have missing variables). These are not design-based in the same sense as the randomization audit because the resampling reference is less well pinned down; we therefore treat them as descriptive early-warning diagnostics that can motivate reweighting, oversampling, or field follow-up.

\paragraph{Selection into the study frame.}
Let \(S_i\in\{0,1\}\) indicate whether unit \(i\) in a broader, policy-relevant universe is enrolled in the experimental sample, or not. When \(\bphi_i\) strongly predicts selection into the study, \(S_i\), the enrolled sample differs systematically from the target universe along pre-treatment features that are visible from space, raising an external-validity warning even if within-sample randomization is sound. We operationalize this as a covariate-shift diagnostic: train a classifier to distinguish enrolled units from units drawn from the putative frame using only \(\bphi\), evaluate strictly out of sample, and summarize fit via likelihood improvement relative to the marginal sampling rate, and permute or randomly re-draw the $S_i$ selection indicator. Because there is no registered or otherwise defensible resampling mechanism for \(S_i\) analogous to \(\Omega\), these quantities are reported as descriptive diagnostics rather than design-valid tests.



\paragraph{Missingness and data quality.}
Let \(R_{ij}\in\{0,1\}\) indicate whether variable \(j\) is observed for unit \(i\). If \(\bphi_i\) predicts \(R_{ij}\) out of sample, then complete-case analyses risk bias because missingness correlates with pre-treatment context that may also shape outcomes, enumerator access, or compliance. We therefore fit response models \(\hat\rho_{ij}(\bphi_i)=\widehat{\Pr}(R_{ij}=1\mid \bphi_i)\) using the same cross-fitting protocol and summarize predictiveness on the likelihood scale relative to the marginal response rate \(\bar{r}_j\), with descriptive randomization inference. A strong signal can help focus field efforts: high-risk or missingness locations can be prioritized for follow-up, instruments can be adapted for hard-to-reach contexts, and data collection modes can be diversified before surveys are fully fielded. Because imagery is strictly pre-treatment and available daily, these diagnostics can be updated in real time during enumeration without peeking at outcomes.

When analysis requires adjustment, the same response models can be pre-specified as building blocks for principled corrections that do not rely on post-treatment information. For variables where missingness is plausibly \emph{at random} given \(\bphi\), inverse-probability weights \(w_{ij}=R_{ij}/\hat\rho_{ij}(\bphi_i)\) or multiple imputation models that condition on \(\bphi\) provide transparent remedies \citep{blackwell2017unified,honaker2011amelia}; doubly robust procedures that combine a response model with an outcome model can be declared in the pre-analysis plan and implemented without altering the design-based logic of the randomization audit \citep{seaman2018introduction}. Where missingness is likely non-ignorable even after conditioning on \(\bphi\), the imagery-based diagnostics still add value by localizing the problem and motivating sensitivity analyses and targeted re-contact. As with selection, we report these quantities as diagnostics rather than as hypothesis tests, and we apply the same cross-validation, sample-splitting, and multiple-testing discipline used elsewhere in the audit.

\begin{table}[!t]
\centering
\small
\begin{tabular}{p{2.7cm} p{4.5cm} p{4.5cm} p{4.5cm}}
\toprule
& \textbf{Randomization audit} & \textbf{Selection audit} & \textbf{Missingness audit} \\
\midrule
\textbf{What it probes} &
Realized assignment vs.\ declared design; pre-treatment covariates; blocks/clusters &
Who/what entered the study vs.\ target population; covariates and inclusion indicators &
Observed vs.\ missing outcomes/units; covariates and missingness indicator \\
\addlinespace
\textbf{Key assumption (focus)} &
Under the registered mechanism, assignment is independent of pre-treatment features (within design) &
Inclusion unrelated to pre-treatment features after stated recruitment rules (diagnostic) &
Missingness unrelated to pre-treatment features (diagnostic) \\
\addlinespace
\textbf{Inferential goal} &
Design-valid check: Is assignment unusually predictable vs.\ the randomization reference? &
Descriptive diagnostic: Is inclusion systematically predictable from features? &
Descriptive diagnostic: Is missingness systematically predictable from features? \\
\addlinespace
\textbf{Typical statistic} &
Predict $A_i$ from $\bM_i$; compare to permutation/CRT under $\Omega$ &
Predict inclusion $S_i$ from $\bM_i$; quantify predictive strength/stability &
Predict missingness $R_{ij}$ from $\bM_i$; quantify predictive strength/stability \\
\addlinespace
\textbf{Primary threat} &
Deviations/manipulations of randomization; hidden stratifications &
Selection bias from sampling/consent/frame &
Attrition/nonresponse distorting the analysis set \\
\addlinespace
\textbf{When to use} &
Pre/post implementation to verify assignment integrity &
During sampling/recruitment; external/internal selection concerns &
During collection/cleaning; nontrivial attrition/nonresponse \\
\addlinespace
\textbf{Actionable follow-ups} &
Re-randomize/re-block; document deviations; sensitivity &
Reweight/adjust recruitment; bounds; document frame &
Weighting/imputation; bounds; follow-up for outcomes \\
\bottomrule
\end{tabular}
\caption{Contrasting randomization, selection, and missingness audits.}
\end{table}

\section{A Preregisterable Workflow}

Having defined the statistic and its design-based calibration, we now turn to practice. The workflow below translates the conditional randomization test into a preregistrable recipe that can be run before, during, or after field mobilization; \emph{mutatis mutandis}, the same steps---swapping the label from assignment to sample membership or response status---produce the selection and missingness diagnostics reported later.


\paragraph{Step 1: Define the design.} Record the experiment's randomization mechanism \(\Omega\): treatment fractions, complete, stratified, or clustered structure, and other constraints. If stratified, list strata membership for each unit. These inputs define the resampling.

\paragraph{Step 2: Acquire strictly pre-treatment imagery.} Select images that unambiguously precede any treatment or mobilization. When archives are sparse or cloudy, use compositing or median mosaics across pre-treatment windows. Avoid sensors whose earliest availability is post-treatment for retrospective analyses (e.g., high-resolution Sentinel 2 data from European Space Agency becomes available only in 2015). Using post-treatment imagery risks mediator/collider bias if later repurposed in outcome models \citep{angrist2009mostly,pearl2009causality,cinelli2024crash}. Archive scene IDs and acquisition dates in the replication package. Clearly define patch size, resolution, bands, normalizations, and features used in processing.


\paragraph{Step 3: Fit the predictive model with sample-splitting.} Use simple learners first (e.g., tree-based models) if the sample size is small; otherwise, consider neural models using satellite imagery or image-derived features to predict treatment. Evaluate out-of-sample (held-out fold or cross-fitting). Save the likelihood-based statistic \(T\).

\paragraph{Step 4: Resample under \(\Omega\) and compute the max-$T$.} Draw \(B\) assignment vectors consistent with \(\Omega\) (e.g., \(B=1{,}000\)), recompute \(T^{(b)}\). If multiple image embedding representations are used, record the maximum test statistic across models per resample. Report adjusted $p$-values and show the observed \(T\) against the reference distribution.

\paragraph{Step 5: Interpret cautiously and report transparently.} A small $p$-value suggests an atypical assignment relative to \(\Omega\), consistent with implementation deviations; a large $p$-value does not prove correct execution, only that the audit detected no image-aligned deviations. Report model choices, pre-treatment windows, resampling details, and multiple-testing adjustments. Provide code and hashes for imagery products to support reproducibility.

\begin{table}[H]
\centering
\begin{tabularx}{\textwidth}{@{}p{1.2cm}p{3.7cm}X@{}}
\toprule
& \textbf{Item} & \textbf{Recommendation} \\ \midrule
$\square$ & Design & Describe \(\Omega\): complete/stratified/clustered; treatment fractions; any constraints.\\
$\square$ & Pre-treatment window & Commit to dates and sensors that strictly precede treatment. Document cloud handling and compositing.\\
$\square$ & Embedding set & Pre-specify models (e.g., CLIP-like, ViT, Swin) and interpretable indices; fix patch size(s) \(s\).\\
$\square$ & Evaluation & Use sample-splitting or cross-fitting; define \(T\) as out-of-sample log-likelihood improvement.\\
$\square$ & Resampling & Set \(B\) (e.g., \(1{,}000\)) and honor blocks/clusters in draws from \(\Omega\).\\
$\square$ & Multiplicity & Use Westfall–Young max-$T$; alternatively, Bonferroni/BH with justification.\\
$\square$ & Outputs & Report adjusted $p$-values, reference distributions, and observed \(T\); archive code and imagery.\\
$\square$ & Auxiliary audits & If used, label as descriptive diagnostics (selection, missingness) and report separately.\\
$\square$ & Ethics \& transparency & Ensure anonymity in replication package as necessitated by ethical standards and required by IRB protocols.\\
\bottomrule
\end{tabularx}
\caption{\textit{Checklist for preregistering and reporting a remote audit.}}
\label{tab:checklist}
\end{table}

\section{Case Study 1: A Remote Audit of the Youth Opportunities Program in Uganda}

We now apply the remote audit to the government-run Youth Opportunities Program (YOP), launched in 2008 in Uganda \citep{blattman2014generating}. Groups of young adults submitted business plans for cash grants; a lottery determined recipients. The trial has been widely cited and influential. We ask whether pre-treatment satellite imagery---without any survey covariates---could have verified randomization and flagged potential issues regarding selection or data missingness early. In this case, we know of no known reports of randomization problems.

\paragraph{Units and imagery.} We treat applicants' villages (geocoded from administrative names) as units. We extract pre-2008 image patches from Landsat archives, which are image composites to mitigate clouds and speckle issues. We compute embeddings from an off-the-shelf backbone used in remote sensing using an EO-fine-tuned CLIP model \citep{li2020rsi}. 
Patch sizes span the village and immediate environs to capture accessibility and settlement structure.

\begin{figure}[htb]
\centering
\includegraphics[width=0.36\textwidth]{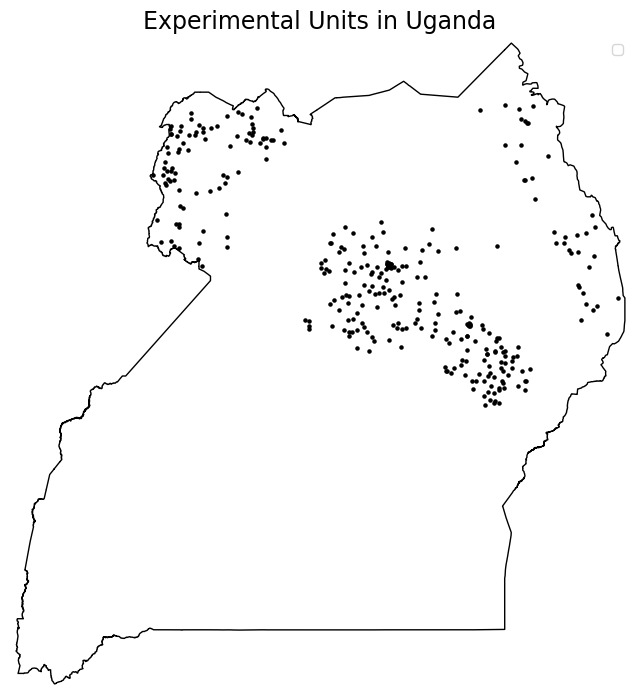}
\caption{\textit{Experimental frame.} Geocoded locations of Youth Opportunities Program (YOP) units in Uganda. The map illustrates nationwide dispersion across settlement types; shaded areas are for orientation only.}
\label{fig:uganda}
\end{figure}

\paragraph{Design and resampling.} We reconstruct the reported randomization scheme (e.g., treatment fractions) from the published record \citep{blattman2014generating}. The CRT resamples treatment vectors consistent with these constraints. For each draw, we recompute the test statistic.

\begin{figure}[htb]
\centering
\begin{subfigure}[b]{0.32\textwidth}
\centering
\includegraphics[width=\textwidth]{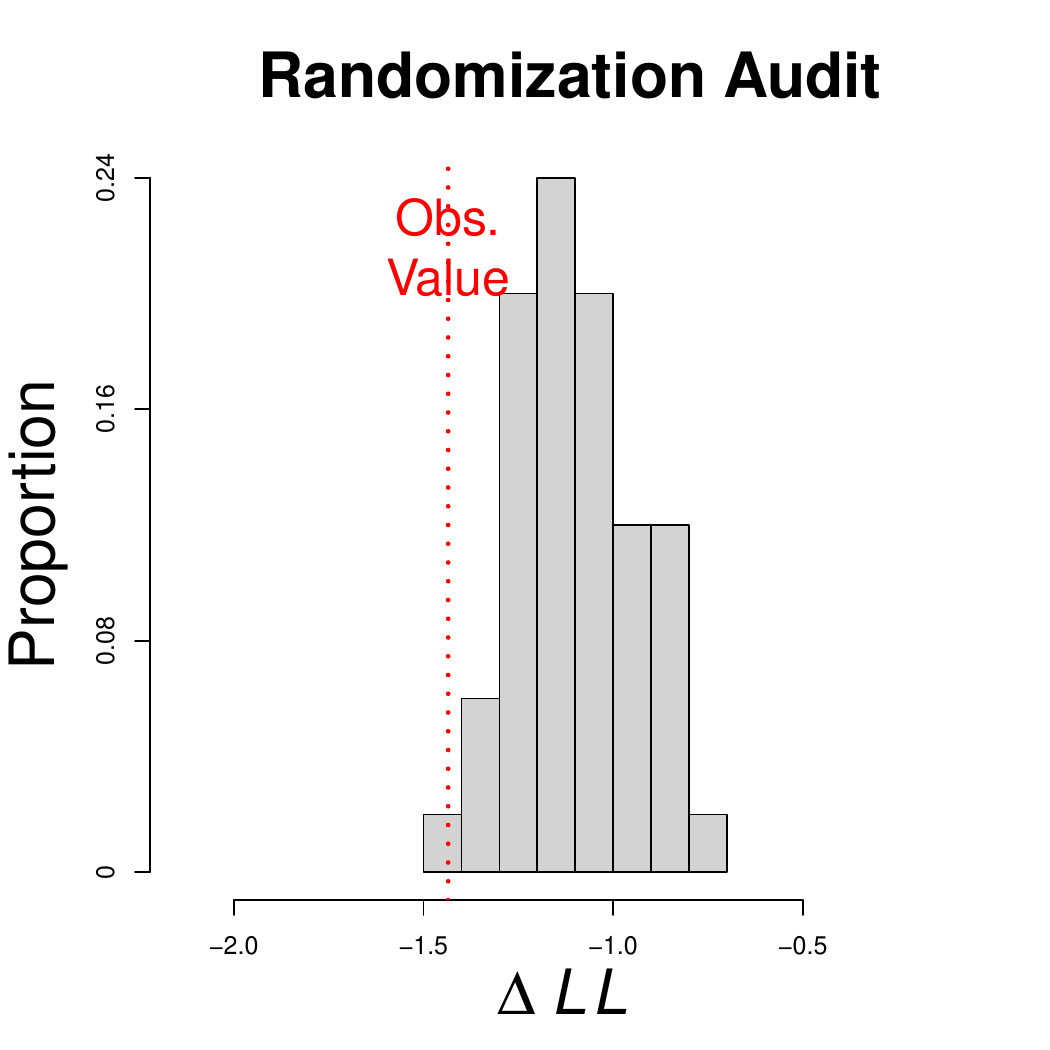}
\caption*{\small Panel A: Randomization audit}
\end{subfigure}\hfill
\begin{subfigure}[b]{0.32\textwidth}
\centering
\includegraphics[width=\textwidth]{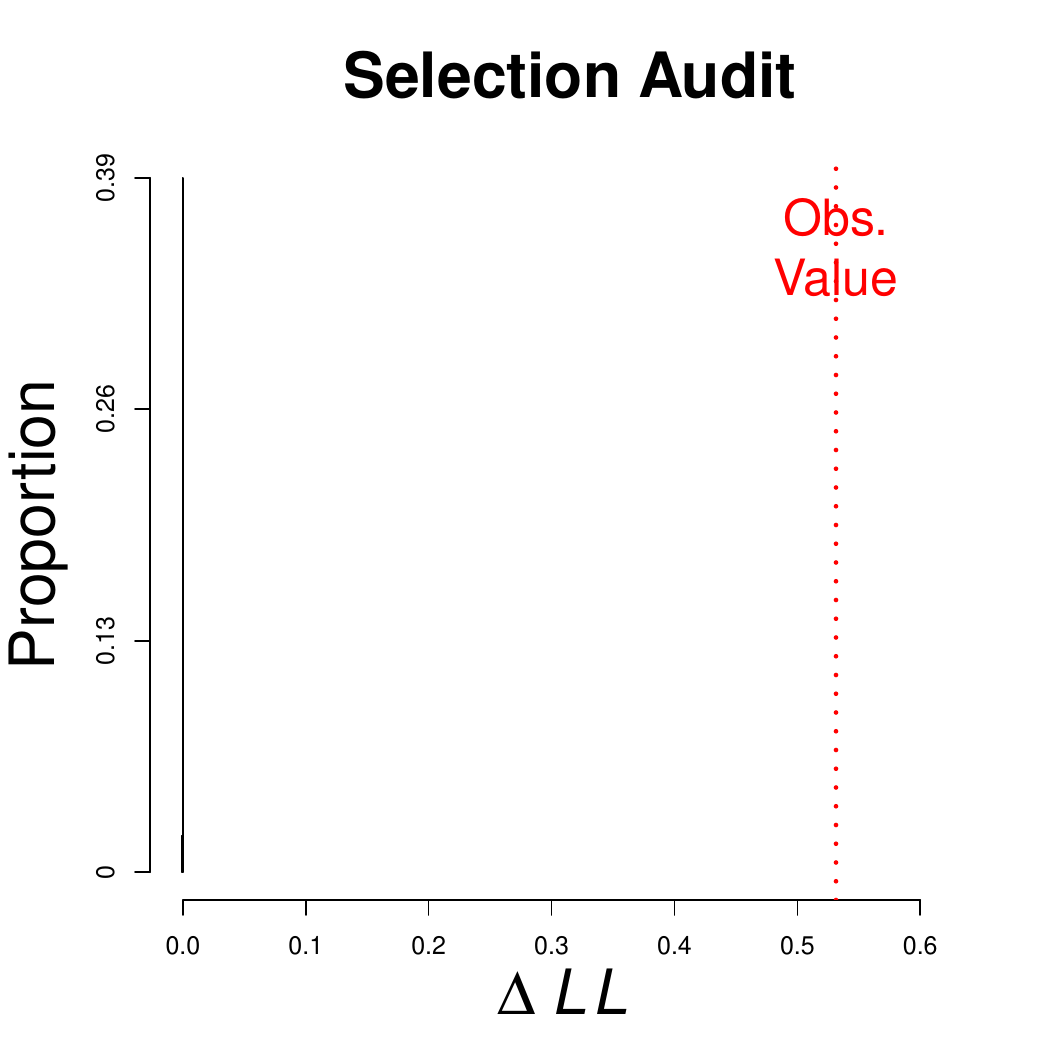}
\caption*{\small Panel B: Selection audit}
\end{subfigure}\hfill
\begin{subfigure}[b]{0.32\textwidth}
\centering
\includegraphics[width=\textwidth]{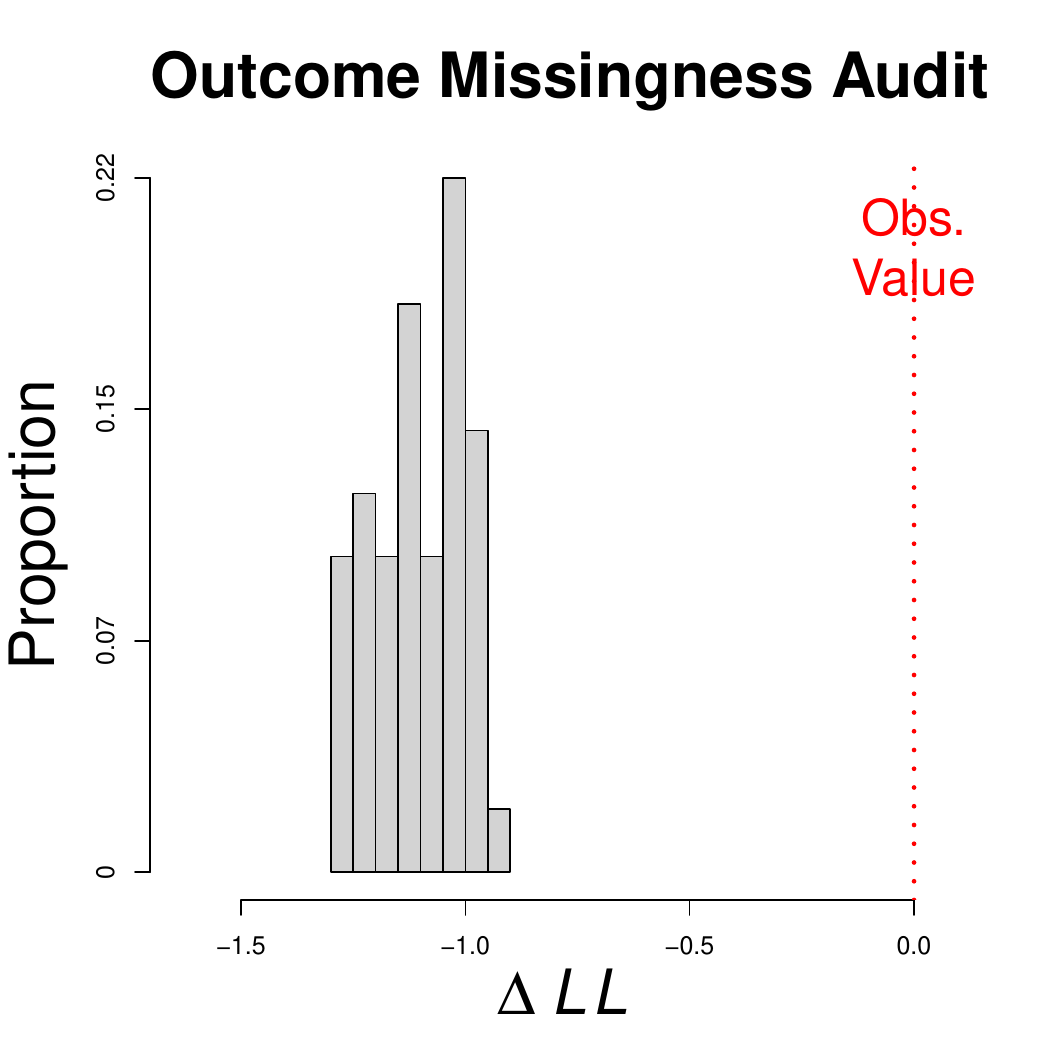}
\caption*{\small Panel C: Missingness audit}
\end{subfigure}
\caption{\textit{Remote audit results.} Each panel displays the reference distribution of the max-statistic obtained from resampling the relevant process (randomization in Panel A; sampling frame or missingness mechanism in Panels B--C) and marks the observed value (vertical line). In Panel A, the observed assignment is not more predictable from imagery than draws from the reported randomization, consistent with integrity of the lottery \citep{blattman2014generating}. Panels B--C highlight auxiliary risks to external validity and systematic missingness.}
\label{fig:panels}
\end{figure}

\paragraph{Results.} Figure~\ref{fig:panels} reports three diagnostics. The \emph{randomization audit} (Panel~A) shows the observed likelihood-improvement statistic (vertical line) within the reference distribution under \(\Omega\); the $p$-value is greater than 0.05, consistent with random assignment not being more predictable from pre-treatment imagery than chance. The two auxiliary checks are here informative. The \emph{selection audit} (Panel~B) contrasts YOP villages against a national frame: image features sharply distinguish enrolled vs.\ unenrolled locations, suggesting external validity concerns if these differences moderate treatment effects \citep{findley2021external}. Finally, the \emph{missingness audit} (Panel~C) indicates that pre-treatment features predict which units have missing variables, suggesting non-random data loss that merits attention in analysis plans (e.g., pre-specified handling of attrition). 

\paragraph{Interpretation and use.} In this application, the remote audit would have (i) corroborated the lottery before expensive baseline enumeration, (ii) offered an early warning about representativeness (selection audit), and (iii) prompted pre-analysis plans for handling missing data differentially by location. None of these conclusions requires survey-collected covariate features. Of course, this audit also does not preclude standard balance tests once baseline data are, in fact, collected; rather, it provides a fast and low-cost early-stage diagnostic that can be embedded in preregistration  \citep{lupia2014openness}.

\section{Case Study 2: Detecting Possibly Faulty Randomization Based on a Retracted RCT in Bangladesh}

\noindent We next apply the remote audit to \citet{begum2022parental}. In this case, the central scientific question is whether the realized assignment conforms to the study's registered mechanism \(\Omega\). We restrict attention here to the design‑valid audit of randomization because independent work by \citet{Bonander2025} has raised concerns about this setting---including evidence that treatment assignment coincided with administrative boundaries and that the assignment vector is identical to one used by some study authors in prior, now‑retracted work \citep{islam2019retracted}. Our conditional randomization test asks a narrower question that fits our framework: is the realized assignment in more predictable from \emph{strictly pre‑treatment local conditions visible in satellite imagery} than would be expected under draws from the purported design \(\Omega\)? The answer is yes, and strongly so---evidence that the observed allocation is atypical of the stated mechanism and consistent with the independent irregularities summarized by \citet{Bonander2025}. Because the audit uses only pre‑treatment imagery and the registered design, it could have been deployed \emph{ex ante} to flag problems with the RCT's implementation before costly on‑the‑ground data collection.


\paragraph{Setup.} Using the replication identifiers from \citet{begum2022parental}, we geolocate \(n\) = \BegumN{} village schools, of which \BegumTreat{} (\BegumTreatSharePct{}\%) are labeled treated. If geolocation is noisy or fails, this would render our tests here conservative, pushing us towards the null hypothesis of independence. Due to the relatively small number of village clusters, we cannot readily deploy large-scale computer vision models (as we could with the larger-scale trial just analyzed, occurring across hundreds of villages). We thus compute low-dimensional, interpretable features from satellite imagery that plausibly reflect long-run local conditions: the median vegetation index (NDVI) and the median nightlight radiance for each unit (median is taken across non-clouded image mosaics from 2008 and 2011, before intervention in 2012). Following the workflow outlined above, we (i) split the sample, (ii) predict treatment from these pre-treatment features using a gradient-boosted tree model (XGBoost), and (iii) summarize the fit with the held-out log-likelihood improvement \(T\) relative to the marginal treatment rate. We then form the finite-sample reference distribution by re-drawing assignments under complete randomization with a fixed treated count \(m\) = \BegumTreat{} (i.e., the \(\Omega\) used here preserves the observed treatment share) and recomputing \(T\) across \(B\) = \BrokenRCTB{} resamples.

\begin{figure}[H]
\centering
\includegraphics[width=0.60\textwidth]{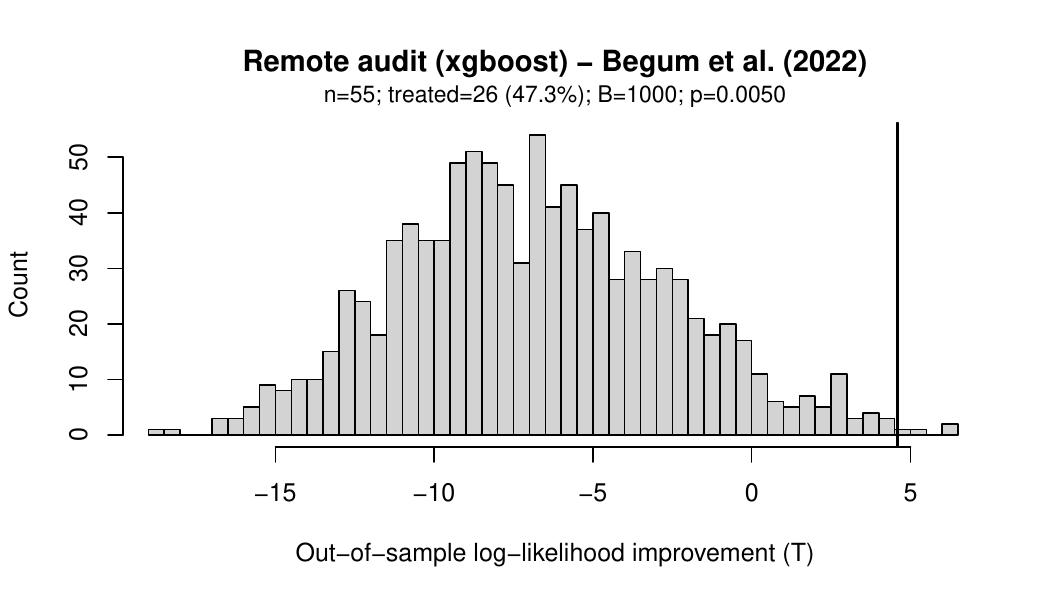}
 \caption{\textit{Remote randomization audit for \citet{begum2022parental} (Cross-fitted XGBoost learner on NDVI and nightlight medians).} Histogram shows the finite-sample reference distribution of the out-of-sample log-likelihood improvement \(T\) under randomization with treated count \(m\) = \BegumTreat{}; the vertical line marks the observed statistic.}
 \label{fig:begum_xgboost}
 \end{figure}

\paragraph{Results.} The XGBoost learners detect assignment predictability that is extreme under \(\Omega\). With a cross-fitted XGBoost tree-based model, the observed improvement falls in the far right tail of the null reference, yielding a design-valid $p$-value of \BegumXgbP{} (Figure~\ref{fig:begum_xgboost}). In words: using only two pre-treatment, physically interpretable proxies of accessibility and local development (greenness and nighttime luminosity), treatment assignment is highly predictable relative to what the reported design would generate by chance. This is precisely the pattern one expects if treatment was targeted to visually distinctive places or if an assignment vector from another exercise was transplanted rather than freshly randomized---concerns documented qualitatively and for related datasets in \citet{begum2022parental}.

\paragraph{Interpretation.} The CRT does \emph{not} identify who or what induced the deviation, nor does it imply that imagery features were used in implementation. It establishes a finite-sample discrepancy: under the claimed design, assignment should not be recoverable from pre-existing landscape signals; yet it is. Coupled with the independent evidence of geographically clustered assignment and shared treatment vectors across linked projects, the audit is consistent with the conclusion that the realized allocation in \citet{begum2022parental} may have deviated from the stated randomization protocol. As with any imagery-only diagnostic, further deviations that are unrelated to what satellites can see remain possible. To conclude, this analysis shows how aspects of randomization integrity can be tested before any fieldwork begins using the remote audit, detecting possible irregularities noted in the replication community. 

\section{Discussion}

Because any remote audit reflects a small number of investigator choices, analysts may naturally ask whether the signal persists under reasonable alternatives. Two quick and informative checks are to vary the strictly proper score (e.g., compare likelihood improvement to Brier improvement) and to assess stability across random seeds and cross-fitting schemes, including spatially robust folds such as leave-one-region-out or leave-one-cluster-out. Reporting the dispersion of the summary statistic $(T)$ across repeats makes the degree of stability transparent.

Scale and sensing choices also matter. Vary image patch sizes to cover plausible geocoding error and local spillovers, and, where multiple pre-treatment sensors exist, consider parallel analyses. Placebo tests that destroy structure, such as remote audits of synthetically allocated treatments, help verify that the pipeline does not manufacture predictability. Complementary stress tests can inject stylized deviations that mimic realistic implementation failures---favoritism toward road-adjacent or administrative-boundary units---to gauge whether the audit would detect such problems at application-relevant sample sizes; a simple, study-calibrated simulation can provide a practical power check.

When the audit suggests potential issues, responses are straightforward. If randomization is flagged, teams might tighten operational constraints or add stratification and re-randomize. They might also strengthen monitoring (seeds, implementation logs) and document deviations. 

If selection is flagged, revisit the sampling frame, clarify the target-population estimand, consider reweighting or targeted enrollment to improve coverage of under-represented areas, and state external validity limits \citep{mullinix2015generalizability}; transported or reweighted estimates may be offered as secondary analyses where appropriate. 

If missingness is flagged, plan additional follow-up in hard-to-reach locations, make small adaptations to instruments or field protocols, and pre-specify imputation or inverse-probability weighting (avoiding complete-case analyses when missingness is predictably related to imagery); attrition bounds provide an additional robustness assessment. Most adjustments are cheaper and cleaner before enumeration begins---the practical advantage of an imagery-only audit is precisely that it surfaces potential problems early enough to improve designs rather than merely document them.

What satellites can test---and what they cannot---is an open and important question. Visual signals reliably register elements of the built and natural environment---settlement structure, roads, roof materials, vegetation, hydrology, and some economic activity such as night lights \citep{henderson2012measuring,jean2016combining,watmough_socioecologically_2019}. These features often co-move with accessibility, wealth, and administrative capacity. A rejection, therefore, indicates an \emph{image-aligned} deviation from $(\Omega)$; it does not identify mechanisms or actors, which require field investigation. Conversely, a non-rejection is not a certificate of perfection: many forces relevant to assignment (patronage networks, norms, internal procedures) leave weak or no satellite trace at available resolutions. The insistence on strictly pre-treatment imagery is central. Using images captured after mobilization risks encoding mediators (construction, publicity) or conditioning on colliders (selection into measurement), with familiar causal consequences. 


Although validity does not depend on interpretability, policy audiences benefit from understanding ``what the model saw.'' Here, comparing interpretable indices---vegetation, built-up, texture---with learned neural network encoders helps facilitate interpretation of how treatment and control differ. Also, post-hoc summaries such as feature importance or representative patch visualizations can aid communication, provided explanation is kept separate from inference. 

In sum, remote audits repurpose broadly accessible pre-treatment imagery to answer a genuinely design-based question---did realized assignment conform to the design? or there residual imbalance, even if the design was faithfully followed?---with a preregistrable, finite-sample-valid procedure that scales across stratified and clustered experiments. They offer a low-cost and fast complement to conventional diagnostics: powerful when deviations align with visible context and deployable early enough to improve designs rather than merely document them.
 
\section{Conclusion}

We develop and demonstrate a \emph{remote audit} of randomization integrity that leverages only pre-treatment satellite imagery and a conditional randomization test. The audit is valid in finite samples, easily preregistered, and compatible with stratified and clustered designs. In Uganda's Youth Opportunities Program \citep{blattman2014generating}, it would have corroborated the reported randomization mechanism while flagging selection and missing-data risks that matter for interpretation and design; in a field experiment in Bangladesh, randomization integrity is itself questioned, in line with concerns raised by an independent team of investigators. 

Remote audits are not a panacea: they detect only image‑aligned deviations, and a large $p$‑value is not a certificate of perfection. Yet the audit's low cost, finite‑sample validity, and preregistrability make it a practically useful tool when ground measurement is slow or difficult. Because the procedure can be run \emph{ex post} as well as \emph{ex ante}, it also enables a broader agenda: a \emph{grand audit} of field‑experimental assignments using only archived pre‑treatment imagery and registered designs. As global archives deepen and off‑the‑shelf vision models improve \citep{li2020rsi,dosovitskiy2020image,liu2021swintransformerhierarchicalvision}, we recommend incorporating remote audits alongside conventional balance checks and process documentation in both preregistration and retrospective quality assurance. \hfill $\square$


\newpage 

\bibliographystyle{apsr}
\bibliography{confoundingbib,referencesAdel}


\clearpage\newpage 

\section*{Appendix}

\medskip
\noindent\textbf{Proposition (Finite-sample validity of the remote audit).}
Fix the pre-treatment embeddings $\bphi=\{\bphi_i\}_{i=1}^n$ and a fold-splitting scheme $H$ (which may be a deterministic function of $\bphi$ or drawn independently of $A$). Let $g(\bphi,A,H)$ be the audit's statistic---e.g., the out-of-sample log-likelihood improvement $T$ computed by training on the $H$-defined training fold and evaluating on the test fold.
Suppose the realized assignment $A$ is drawn from the registered randomization mechanism $\Omega$ and is (by design) independent of all pre-treatment variables, including $\bphi$.
For $b=1,\ldots,B$, draw $A^{(b)}\sim\Omega$ (independently of each other and of $A$), and define $T=g(\bphi,A,H)$ and $T^{(b)}=g(\bphi,A^{(b)},H)$, recomputing the learner under each $A^{(b)}$.
Then the $p$-value
\[
p=\frac{1}{B+1}\Big(1+\sum_{b=1}^B \1\{T^{(b)}\ge T\}\Big)
\]
satisfies $\Pr\!\big(p\le\alpha\mid \bphi,H\big)\le \alpha$ for all $\alpha\in[0,1]$.
Hence the remote audit controls Type~I error at level $\alpha$ in finite samples, conditional on $(\bphi,H)$.
The result continues to hold for stratified or clustered designs provided $\Omega$ and the resampling preserve the design's block/cluster constraints.

\emph{Proof.} 
Condition on $(\bphi,H)$.
Under the null, $A\sim\Omega$ and $A^{(1)},\ldots,A^{(B)}\stackrel{\text{i.i.d.}}{\sim}\Omega$ are exchangeable.
Applying the fixed, measurable map $g(\cdot)$ to each assignment yields exchangeable statistics $\big(T,T^{(1)},\ldots,T^{(B)}\big)$.
Therefore the rank of $T$ among these $B{+}1$ values is uniformly distributed on $\{1,\ldots,B{+}1\}$ (with ties handled by the $\ge$ rule or broken at random), which implies that $p$ is (super-)uniform and $\Pr(p\le\alpha\mid \bphi,H)\le \alpha$.
When $\Omega$ imposes block or cluster totals, exchangeability holds conditional on those totals, so the same argument applies. \hfill $\square$

\medskip
\noindent\textit{Remark.}
Refitting the predictive model within each resample is what ensures that $g(\bphi,\cdot,H)$ treats every draw from $\Omega$ symmetrically; reusing a learner trained only on the realized $A$ can break exchangeability. See \citet{candes2018panning} and \citet{hennessy2016conditional} for more information.

\end{document}